\documentclass[journal=jacsat,manuscript=communication]{achemso}

\usepackage[version=3]{mhchem} 

\usepackage{hyperref}
\usepackage{graphicx}
\usepackage{amsmath}
\usepackage{physics}
\usepackage[caption=false]{subfig}
\hypersetup{
	colorlinks=true,       
	linkcolor=blue,        
	citecolor=blue,        
	filecolor=magenta,     
	urlcolor=blue         
}
\affiliation{Department of Chemistry, Purdue University, West Lafayette, IN 47907, USA}
\author{Sumit Suresh Kale}
\affiliation{Department of Chemistry, Purdue University, West Lafayette, IN 47907, USA}
\author{Sabre Kais}
\email{kais@purdue.edu}
\affiliation{Department of Chemistry, Purdue University, West Lafayette, IN 47907, USA}
\altaffiliation{Department of Electrical and Computer Engineering, Purdue University, West Lafayette, IN 47907, USA}
\altaffiliation{Purdue Quantum Science and Engineering Institute, West Lafayette, IN 47907, USA}
\email{kais@purdue.edu}
\title[An \textsf{achemso} demo]
  {Simulation of Chemical Reactions on a Quantum Computer}

\abbreviations{PA,SOC,BEC}

\begin{document}
\begin{abstract}
Studying chemical reactions, particularly in the gas phase, relies heavily on computing scattering matrix elements. These elements are essential for characterizing molecular reactions and accurately determining reaction probabilities. However, the intricate nature of quantum interactions poses challenges, necessitating the use of advanced mathematical models and computational approaches to tackle the inherent complexities. In this study we develop  and apply  a quantum computing algorithm for the calculation of scattering matrix elements. In our approach we employ the time dependent method based on the M$o\llap{/}$ller operator formulation where the S-matrix element between the respective reactant and product channels is determined through the time correlation function of the reactant and product M$o\llap{/}$ller wavepackets. We successfully apply our quantum algorithm to calculate scattering matrix elements for 1D semi-infinite square well potential and on the co-linear hydrogen exchange reaction. As we navigate the complexities of quantum interactions, this quantum algorithm is general and emerges as a promising avenue, shedding light on new possibilities for simulating chemical reactions on quantum computers. 
\end{abstract}

\section{Introduction}
In recent years, there have been significant advancements in the field of quantum information and quantum computing. Both hardware and software have progressed rapidly, leading to the development of algorithms for the current quantum computers. These algorithms show promising results in solving research challenges that are beyond the capabilities of even the most powerful conventional supercomputers \cite{arute2019quantum,kim2023evidence}. Central to the efficacy of such algorithms exists fundamental quantum properties, including superposition, entanglement, coherence, and interference. Recent advancements in quantum hardware have spurred a swift surge in the creation of innovative quantum algorithms\cite{bharti2022noisy,sajjan2022quantum}. The predominant focus of algorithmic development encompasses a diverse array of topics such as spectroscopy\cite{sawaya2021near,bruschi2024quantum,johri2017entanglement}, electronic structure\cite{xia2017electronic,xia2018quantum,sajjan2023physics}, vibrational structure\cite{ollitrault2020hardware}, quantum many-body problems\cite{bravyi2019approximation}, and open quantum dynamics\cite{hu2020quantum}. Notably, the application of these algorithms to address scattering problems has received limited attention thus far. This study aims to bridge this gap by developing a quantum algorithm specifically designed to estimate scattering matrix elements for both elastic and inelastic collision processes.

Quantum scattering calculations play a crucial role in advancing our understanding of fundamental physical and chemical phenomena\cite{althorpe2003quantum}, making them highly significant across diverse scientific disciplines, including the study of chemical reaction mechanisms in the gaseous phase\cite{fu2017recent} and atmospheric chemistry\cite{guitou2015quantum,madronich1999role,Levine}. These calculations are indispensable for the accurate interpretation of experimental findings in gas-phase interactions, providing intricate insights into bimolecular chemical reactions\cite{zhang2016recent}. Moreover, quantum scattering theory serves as a valuable tool for calculating essential parameters like cross-section and reaction rate in atomic, and chemical physics, contributing to the study of many scattering processes\cite{pozdneev2019application}. The necessity of full-dimensionality in low-energy molecular scattering calculations underscores their pivotal role in comprehensively understanding complex molecular interactions. Additionally, molecular scattering experiments, particularly those conducted under cold\cite{krems2008cold} and ultracold\cite{mellish2007quantum} conditions, yield unparalleled insights into intermolecular interactions. A comprehensive understanding of the quantum dynamics in ultracold environments has the potential to reveal innovative strategies, possibly employing quantum phenomena such as superposition, entanglement, and interference patterns to control reaction outcomes\cite{shapiro1997quantum,kale2021constructive,aoiz2018quantum}.

There are two ways to approach the quantum scattering problem in chemical reactions one can employ either the Time Independent (TI) or the Time Dependent (TD) formalism. Quantitatively accurate simulations of quantum scattering, achieved by directly solving the TI Schrödinger equation through methods like coupled-channel techniques\cite{althorpe2003quantum,zhang2016recent} or basis set expansion, represents as a computational bechmark. Various techniques in the time-independent formulation have been developed to address this issue, such as the S-matrix version of the Kohn variational principle. This involves applying a variational principle to determine expansion coefficients in scattering coordinates, resulting in a more efficient and practical approach for quantum scattering calculations. But in general the TI approach suffers from the classic "curse of dimensionality" problem meaning the computational scaling scales exponentially as the problem size. 

To mitigate this challenge, numerous quantum algorithms have been proposed in the literature for different applications\cite{bian2021quantum,kassal2008polynomial,du2021quantum}, with many relying on the Quantum Phase Estimation (QPE) algorithm. However, as QPE is a fault-tolerant algorithm, its implementation is currently unfeasible on utility-scale quantum processors. Consequently, its application is constrained to smaller-scale problem sets. Recently, Xing et al. \cite{xing2023hybrid} proposed an innovative solution by employing the S-matrix version of the Kohn variational principle to address the scattering problem. This approach alleviates the intricate task of symmetric matrix inversion through the utilization of the Variational Quantum Linear Solver (VQLS)\cite{bravo2023variational}. Nonetheless, there is a need for further improvement in the scalability of VQLS on Noisy Intermediate-Scale Quantum (NISQ) devices and the trainability of the variational ansatz, especially when addressing larger and more complex problems. In TD formalism of quantum scattering problem wavepackets are constructed and allowed to evolve using the time-dependent Schrödinger equation. TD methods present the ability to extract information across a range of translational energies in a single computational iteration and exhibit superior scalability compared to their  TI counterparts. Furthermore, TD methods offer a more comprehensive understanding of dynamic processes, enabling the investigation of time-evolution phenomena in scattering events. This aspect becomes particularly crucial for reactions involving the formation of new chemical bonds. Despite these potential advantages, as of our current knowledge, there is no existing quantum algorithm based on the TD formalism\cite{das1990time} for the scattering problem.

Here we propose a TD Quantum Algorithm based on the M$o\llap{/}$ller operator formulation of the S-matrix\cite{weeks1993time,tannor1993wave}. The fundamental technical procedure involves dynamics\cite{kosloff1988time,agmon1987dynamics} of two wavepackets: one representing an asymptotic reactant localized in a single reactant channel, and the other representing an asymptotic product localized in a single product channel. The selection of specific reactant and product channels dictates the S-matrix elements to be computed. These wavepackets are then advanced towards the interaction region, one forwards (reactant) and the other backwards (Product) in time. Subsequently, the resulting reactant and product wavepackets are transformed into a common representation\cite{zhang1990new,das1990time}, and the correlation function $ C_{\gamma',\gamma}(t)= \bra{\Psi^{\gamma'}_{-}}\exp{-iHt}\ket{\Psi^{\gamma}_{+}}$ is calculated between their subsequent time-dependent evolution. Where, $\ket{\Psi^{\gamma'}_{-}}$ and $\ket{\Psi^{\gamma}_{+}}$  corresponds to the product and reactant M$o\llap{/}$ller states in the $\gamma'$ and $\gamma$ channel, and $H$ represents the total hamiltonian. Calculation of the correlation function  is the core computational capability of the proposed quantum algorithm and we employ modified version of the Hadamard test to estimate the correlation function. The M$o\llap{/}$ller operator formalism is then employed to express S-matrix elements between the chosen reactant and product channels in terms of the Fourier transform of the correlation function computed. This expression enables the computational effort to be efficiently directed towards computing only the S-matrix elements that are of interest. We discuss the theoretical and computations details in the further sections.

\section*{Time Dependent Formulation of the Scattering Matrix}\label{td_theory}
\textbf{\textbf{Time Dependent Formulation of the Scattering Matrix}}

Asymptotic reactant state in the $\gamma^{th}$ arrangement channel is represented by $\ket{\Psi^{\gamma}_{in}}$ and the channel Hamiltonian $H^{\gamma}_{o}$ governs the asymptotic dynamics of the reactant state wavepacket. Assuming $H^{\gamma}_{o}$ is time independent evolution of the reactant state wavepacket is given by:\footnote{In all the derivations we assume atomic units}

\begin{equation}
       \ket{\Psi^{\gamma}_{in}(x_{\gamma},t)}  = \exp{-iH^{\gamma}_{o}t} \ket{\Psi^{\gamma}_{in}(x_{\gamma},0)} 
\end{equation}

The wavefunction $\ket{\Psi^{\gamma}_{in}(x_{\gamma},t)}$ can be represented in the $\ket{x_{\gamma}} = \ket{r_{\gamma}}\ket{R_{\gamma}}$ coordinate representation where $\ket{r_{\gamma}}$ and $\ket{R_{\gamma}}$ represents internal quantum state and relative position of reactants respectively.  Additionally the center of mass (COM) is stationary in the coordinate representation and the asymptotic Hamltonian is separable in two parts $H^{\gamma}_{o} = H^{\gamma}_{rel} + H^{\gamma}_{int}$, here
$H^{\gamma}_{int}$ and $H^{\gamma}_{rel}$ corresponds to Hamiltonians which governs the internal dynamics and relative motion respectively. Eigenfunctions of $H^{\gamma}_{rel}$ $(\ket{ k_{\gamma}})$ and $H^{\gamma}_{int}$ $(\ket{\gamma})$ spans the $\gamma^{th}$ channel momentum representation. Thus the $\gamma^{th}$ channel momentum representation is given by: $\ket{k_{\gamma},\gamma} = \ket{ k_{\gamma}} \ket{\gamma}$ and the the incoming wavepacket is conveniently expressed in the momentum representation. Similarly, an asymptotic product state 
$\ket{\Psi^{\gamma'}_{out}}$ belongs to the $\gamma'$th channel Hilbert space. In this scenario, the channel label $\gamma'$ specifies the product arrangement channel along with all the internal quantum numbers of the products. The asymptotic dynamics of the product wavepacket is entirely governed by the $\gamma'$th channel Hamiltonian $H^{\gamma'}_{o}$. The asymptotic Hilbert space is constructed as the direct sum of all individual channel Hilbert spaces, playing a pivotal role in achieving a complete representation of the scattering operator $S$. 

Near the interaction region the dynamics of the wavepacket localized in $\gamma$th channel is controlled by the complete Hamiltonian $H = H^{\gamma}_{o}+V$, where $V$ denotes the interaction potential. The isometric M$o\llap{/}$ller opeator is defined as:

\begin{equation}
    \Omega^{\gamma}_{\pm} = \lim_{t\rightarrow \mp \infty}\left[ \exp{(iHt)} \exp{(-iH^{\gamma}_{0}t)} \right] 
    \label{moller_op}
\end{equation}
M$o\llap{/}$ller operator mentioned in Eq. \ref{moller_op} is fundamentally a composition of time evolution operators, with one corresponding to the asymptotic Hamiltonian of the $\gamma$th channel $H^{\gamma}_{o}$ and the complete Hamiltonian $H$. Given an asymptotic state $\ket{\Psi^{\gamma}_{in(out)}}$ at time $t=0$ the impact of applying M$o\llap{/}$ller $\Omega^{\gamma}_{+(-)}$ involves backward(forward) propagation to time $t = \tau = -\infty (+\infty)$ under the asymptotic Hamiltonian $H^{\gamma}_{o}$ followed by forward (backward) propagation under the full Hamiltonian $H$ from $t=-\tau (+\tau)$ to $t=0$. In the coordinate representation the asymptotic time limit $\tau$ is the time required to propagate wavefunction from the interaction region to the asymptotic region\cite{engel1988relative}. In brief the impact of M$o\llap{/}$ller operator on  $\ket{k_{\gamma},\gamma}$ is to create a new set of states $\ket{k_{\gamma},\gamma\pm}$

\begin{equation}
   \ket{\Psi^{\gamma}_{\pm}} =  \lim_{t\rightarrow \mp \infty}\left[ \exp{(iHt)} \exp{(-iH^{\gamma}_{0}t)} \right] \ket{\Psi^{\gamma}_{in/out}}
   \label{moller_states}
\end{equation}

\begin{equation}
    \ket{\Psi^{\gamma}_{in(out)}} = \int_{-\infty}^{+\infty} dk_{\gamma} \eta_{\pm}(k_{\gamma}) \ket{k_{\gamma},\gamma}
    \label{psi_k}
\end{equation}

While computing the $S$ matrix elements in the momentum representation it's crucial to note an important property of the basis vectors of the $\gamma$th channel. These basis vectors forms an eigen-basis for their respective asymptotic Hamiltonians $H^{\gamma}_{o}$ and the eigenvalues corresponding to $\ket{k_{\gamma},\gamma}$ is $\frac{k^{2}_{\gamma}}{2}+E_{\gamma}$. Here $\frac{k^{2}_{\gamma}}{2}$ corresponds to the relative kinetic energy and $E_{\gamma}$ relates to the internal energy. 
One can use the intertwining relation $\Omega^{\gamma}_{\pm} H^{\gamma}_{o}=H\Omega^{\gamma}_{\pm}$ to show that the $\ket{k_{\gamma},\pm\gamma}$ basis set forms an eigen-basis for $H$ with the eigenvalues that corresponds to $\ket{k_{\gamma},\gamma}$\cite{weeks1993time}. The correlation function between the M$o\llap{/}$ller states is defined in Eq. \ref{eq_correl_fun}. One can obtain the correlation function using two different approaches, in the first approach $\ket{\Psi^{\gamma}_{+}}$ is propagated under the full Hamiltonian $H$ from time $t=0$ to $t=\tau$ and take an inner product of the resulting propogated wavepacket with the product M$o\llap{/}$ller state $\ket{\Psi^{\gamma'}_{-}}$. The an alternative approach the time evolution operator $(\exp{-iHt})$ can be applied symmetrically, first $\bra{\Psi^{\gamma'}_{-}}$ is propagated from $t=0$ to $t=-\tau$ and then $\ket{\Psi^{\gamma}_{+}}$ from $t=-\tau$ to $t=\tau$\cite{tannor1993wave}. The scattering matrix elements can be expressed in terms of Fourier transform of the calculated correlation function (discussed further in the Supplementary Information). 

\begin{equation}
    C_{\gamma',\gamma}(t) = \bra{\Psi^{\gamma'}_{-}}\exp{(-iHt)}\ket{\Psi^{\gamma}_{+}}
    \label{eq_correl_fun}
\end{equation}

\pagebreak

\section*{Quantum Circuit implementation}\label{quant_circuit}
\textbf{Quantum Circuit implementation}

The quantum algorithm focuses on computing the correlation function $C_{\gamma',\gamma}$ described in Eq. \ref{eq_correl_fun}. While the conventional Hadamard test, can determine the expectation value of an operator $U$ defined as $\bra{\Psi}U\ket{\Psi}$, it is not capable of calculating the correlation function. The reason behind this limitation is that the correlation function, unlike a straightforward expectation value, involves a computation similar to assessing the overlap $(\langle\Psi_{1}|\Psi_{2}\rangle)$ between two quantum states, $ \ket{\Psi_{1}}$ and $\ket{\Psi_{2}}$. Where $\ket{\Psi_{1}} = \ket{\Psi^{\gamma'}_{-}}$ and  $\ket{\Psi_{2}} = \exp{(-iHt)} \ket{\Psi^{\gamma}_{+}}$. To calculate the state overlap $(\langle\Psi_{1}|\Psi_{2}\rangle)$ a slight modification is required in the conventional Hadamard test. We define operators $O_{\gamma}$ ($O_{\gamma'}$)  such that when we operate them on the vacuum state $(\ket{000\ldots})$ they prepare $\ket{\Psi_{1}}$ $(\ket{\Psi_{2}})$ quantum states.  

\begin{figure*}[!ht]
    \centering
    \includegraphics[width=0.8\linewidth]{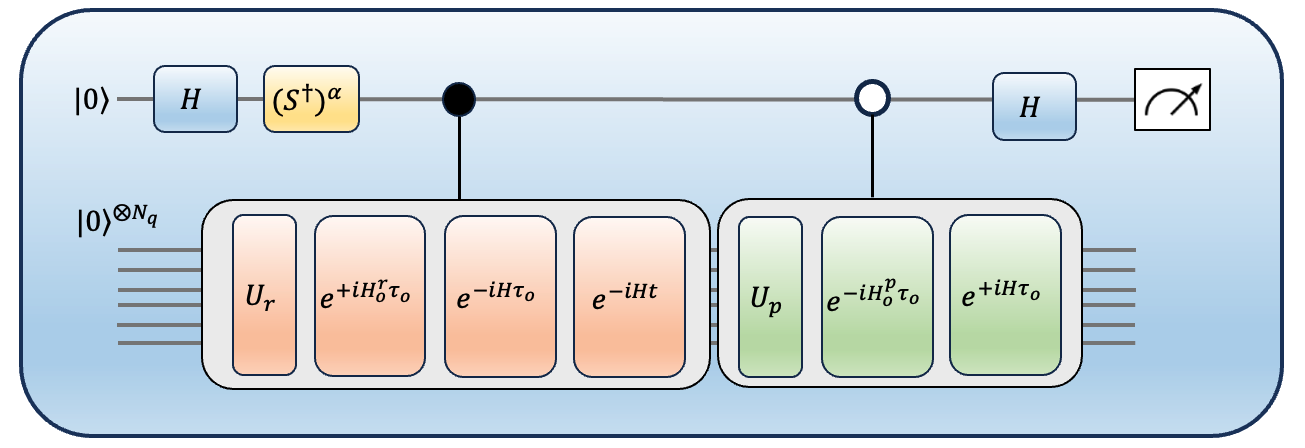}
    \caption{Quantum circuit that calculates the correlation function $C_{\gamma',\gamma}(t)$ defined in Eq. \protect\ref{eq_correl_fun}. The circuit includes two quantum registers a single qubit ancillary register which is measured and stores information about the correlation function and $N_{q}$ qubit quantum register that encodes and manipulates the quantum state. }. 
    \label{fig:corr_hada_circ_new}
\end{figure*}

\begin{equation}
    O_{\gamma}(t) = e^{-iHt} e^{-iH\tau_{o}} e^{+iH^{\gamma}_{o}\tau_{o}} U_{r}
    \label{op_og}
\end{equation}

\begin{equation}
    O_{\gamma'} = e^{+iH\tau_{o}} e^{-iH^{\gamma'}_{o}\tau_{o}} U_{p}
    \label{op_ogdg}
\end{equation}

In Eq. \ref{op_og} operator $U_{r}$ corresponds to a state preparation that transforms the initial vacuum state into the asymptotic reactant channel wavepacket $\ket{\Psi^{\gamma}_{in}}$ similarly $U_{p}$ in Eq. \ref{op_ogdg} is defines such that:  $\ket{\Psi^{\gamma'}_{out}} = U_{p}\ket{000\ldots}$. 
Similar to the Hadamard test here we have two quantum registers Ancillary qubit $q_{a}$ that stores the information about the correlation function and qubit $q_{s}$ that stores and manipulates the quantum state. We first apply Hadamard operation on the ancillary qubits (followed by a phase gate $S^{\dagger}$ if the objective is to calculate $Im(C_{\gamma',\gamma}(t))$). Next we apply two controlled operations which are at this algorithm's core, as shown in Figure \ref{fig:corr_hada_circ_new}. In the first controlled operation we operate the controlled version of operator $O_{\gamma}$ Eq. \ref{op_og} with ancillary qubit $q_{a}$ as a control and the the state qubit $q_{s}$ as a target with the control state '$\ket{1}_{q_{a}}$'. Meaning that the operator $O_{\gamma}$ will operate on the the state quibts $q_{a}$ only when the quantum state of the ancillary qubit $q_{a}$ is $\ket{1}$. Similarly we apply another controlled version of the $O_{\gamma'}$ operator (Eq. \ref{op_ogdg}) with the ancillary qubit $q_{a}$ as control and the control state '$\ket{0}_{q_{a}}$'. Finally we apply the Hadamard operation and measure the ancillary qubit is measured in the $\sigma_{z}$ basis.

The wavefunction is expressed in the momentum representation Eq. \ref{psi_k} on $q_{s}$ and binary encoding\cite{sawaya2020resource} is used to map it to qubit wavefunction. The wavepacket propagation including the Hamiltonian encoding is done in first quantization by expressing the asympotic $H^{\gamma}_{o}$ and total Hamiltonian $H$ in the momentum representation basis and then expressing the Hamiltonian as a linear combination of Pauli strings. The propagators $\exp{(\pm iHt)}$ can be approximated by employing higher order Trotter-Suzuki decompositions\cite{hatano2005finding}. One can use the Qiskit\cite{mckay2018qiskit} Open Source SDK to simulate these quantum circuits (discussed further in the Supplementary Information) .

\section*{Applications}\label{application}
\textbf{Applications}

\subsection{A One-dimensional Semi-infinite Well}\label{'1d_example}
\textbf{A One-dimensional Semi-infinite Well}

\begin{figure}[!ht]
    \centering
    \includegraphics[width=0.85\linewidth]{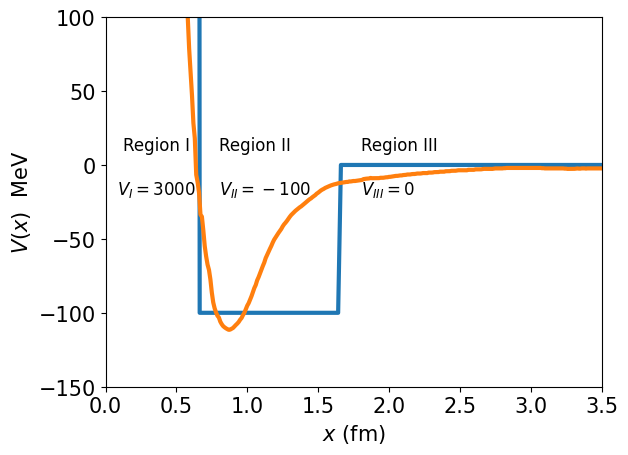}
    \caption{Semi-infinite square well approximation of the $^{1}S_{0}$ potential (Orange Curve). The potential is divided in three different regions, Region I $(x\leq 0.65 fm)$, Region II $(0.65< x \leq 1.65)$ and Region III $x\geq1.65 fm$. } 
    \label{fig:1D_potential_figure}
\end{figure}

Evaluating the effectiveness of a new technique often involves applying it to a problem with a well-established analytical solution. In this section, we employ the proposed quantum algorithm to determine the scattering matrix element for a two-nucleon scattering problem. Previous studies\cite{davis2010time} 
have demonstrated that the widely utilized $^{1}S_{0}$ Argonne V18 (AV18) potential\cite{wiringa1995accurate}  which plays a crucial role in describing nucleon-nucleon interactions and is among the most commonly used potentials, can be effectively approximated by a semi-infinite square well with dimensions resembling those of the original potential as shown in Fig: \ref{fig:1D_potential_figure}.

\[
V(x) =
\begin{cases}
  3000 \quad MeV & \text{if } \hspace{0.2cm} x \leq 0.65 fm \\
  -100 \quad MeV & \text{if } \hspace{0.2cm}  0.65  fm< x \leq 1.65  fm \\
  \hspace{0.1cm} 0 \quad MeV & \text{if } \hspace{0.2cm} x > 1.65  fm
  \label{eq:1d_poten}
\end{cases}
\]

The functional form of the reactant and product wavepackets at time $t=0$ in the co-ordinate representation is given in Eq. \ref{eq:1d_wfn_x_t=0}. The reactant/product wavepackets are defined as a Gaussian centered at $x=x_{o}$ with the spread of $\Delta x_{o}$ and travelling with the momentum of $k_{o}$. The same wavefunction can be expressed in the momentum representation as shown in Eq. \ref{eq:1d_wfn_k_t=0}. From Heisenberg's uncertainty principle it's clear that increase in the $\Delta x_{o}$ spread in the co-ordinate space results in decrease in the momentum representation and vise versa.

\begin{equation}
    \Psi_{in}(x,t=0) = \left( \frac{1}{2\pi\Delta x^{2}_{0}} \right)^{\frac{1}{4}} \exp{ \left[\frac{-(x-x_{o})^{2}}{4\Delta x^{2}_{o}} +ik_{o} (x-x_{o}) \right]} 
    \label{eq:1d_wfn_x_t=0}
\end{equation}

\begin{equation}
    \Psi_{in}(k,t=0) = \left( \frac{2\Delta x^{2}_{0}}{\pi} \right)^{\frac{1}{4}} \exp{ \left[ -\Delta x^{2}_{o}(k-k_{o})^{2}+ix_{o}k \right]} 
    \label{eq:1d_wfn_k_t=0}
\end{equation}

\begin{figure}[h!]
    \centering
    \includegraphics[width=0.75\linewidth]{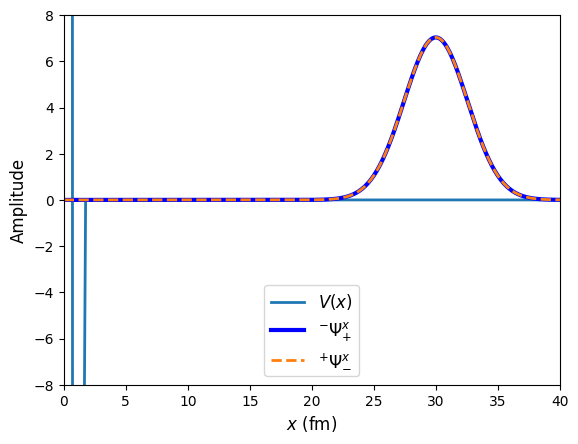}
    \caption{Absolute value of the M$o\llap{/}$ller states $\ket{^{\mp}\Psi^{x}_{\pm}})$ in the position representation Eq. \protect\ref{eq:1d_wfn_x_t=0}. The potential $V(x)$ is also shown in solid Azure curve. The potential Absolute values of the wavepacket are scaled by the factor of 0.1 and 15 respectively.}
    \label{fig:1d_wfn_x_t=0}
\end{figure}

\begin{figure}[h!]
    \centering
    \includegraphics[width=0.75\linewidth]{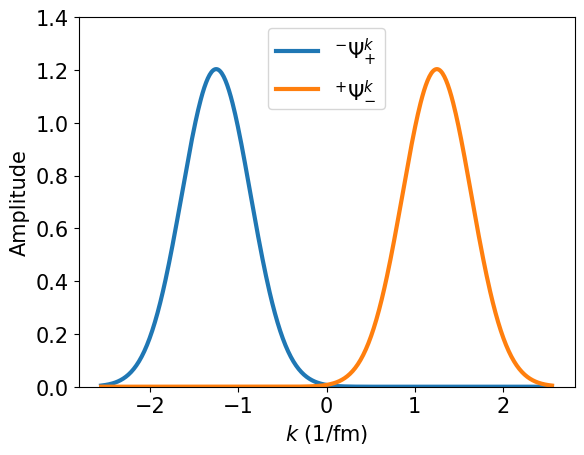}
    \caption{Absolute value of the reactant ($\ket{^{-}\Psi^{k}_{+}})$) and product $\ket{^{+}\Psi^{k}_{-}})$ M$o\llap{/}$ller states  in the momentum representation Eq. \protect\ref{eq:1d_wfn_k_t=0}.}
    \label{fig:1d_wfn_k_t=0}
\end{figure}

Parameters are chosen carefully for Eq. \ref{eq:1d_wfn_k_t=0} such that the wavepacket should only contain either positive or negative contribution of the momentum else the wavepacket splits up (For more information please refer to the Supplementary Information). Fig. \ref{fig:1d_wfn_x_t=0} plots the scaled potential $V(x)$ and the amplitude of the reactant and product M$o\llap{/}$ller states in the position representation. Fig. \ref{fig:1d_wfn_x_t=0} shows that none of the wave-packet is present within the potential, so there is no need to perform the initial propagation of the wave-packets to the asymptotic limit under the channel asymptotic Hamiltonian and subsequent back propagation under the full channel Hamiltonian (Eq. \ref{moller_op}). Thus in this specific scenario, these states already correspond to the M$o\llap{/}$ller states as outlined in the Eq.\ref{moller_states}. Fig. \ref{fig:1d_wfn_k_t=0} plots the amplitude of the reactant (Azure) and product(Orange) M$o\llap{/}$ller states in the momentum representation. In the legend $^{-}\Psi^{k}_{+}$ subscript $+ (-)$ corresponds to reactant(product) M$o\llap{/}$ller state and the superscript $+ (-)$ corresponds to the sign of contributing $k$ values. The reactant M$o\llap{/}$ller state has contributions from negative values of $k$ and would propagate towards the interaction region and interact with the semi-infinite well.

Referring back to the TD theory of reactive scattering discussed in section \ref{td_theory} we clearly see that there is no internal degree of freedom and we can represent the initial wavepackets Eq. \ref{eq:1d_wfn_k_t=0} in the plane wave basis and get the $\eta_{\pm}(k)$ values and the Hamiltonian matrix $(H = T+ V)$ which can be expressed as a linear combination of Pauli matrices. Since we are working with the plane wave basis the Kinetic Energy (KE) $T$ matrix will be diagonal and easier to impliment. We discretize the $x$ and $k$ space in 256 grid points, since we encode the initial wavefunciton using binary encoding this problem requires 8 qubits ($log_{2}(256)$) to encode the wavefunction. Since we are already starting from the reactant and product M$o\llap{/}$ller states the operators $O_{\gamma}(t)$ and $O_{\gamma'}$  defined in Eq. \ref{op_og} and Eq. \ref{op_ogdg} reduces down to $e^{-iHt}U_{r}$ and $U_{p}$ respectively.

\begin{figure*}[t]
    \centering
    \includegraphics[width=0.85\linewidth]{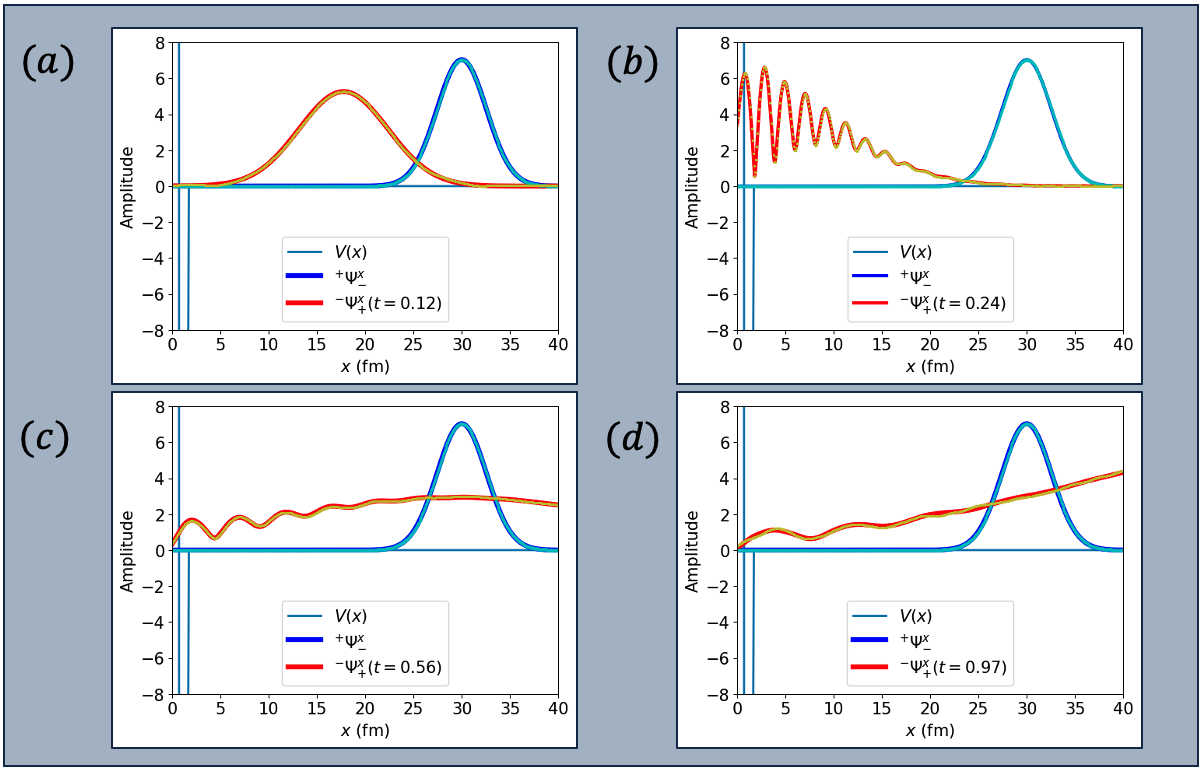}
    \caption{Scaled absolute value of the M$o\llap{/}$ller states $\ket{^{\mp}\Psi^{x}_{\pm}})$ in the position representation Eq. \protect\ref{eq:1d_wfn_x_t=0}. The potential $V(x)$ is also shown in solid Azure curve. Blue curve: product M$o\llap{/}$ller state, Red curve: propagated reactant M$o\llap{/}$ller state at time (a) $t=0.12$,(b) $t=0.24$,(a) $t=0.56$ and (a) $t=0.97$. Yellow circles: Quantum simulation of the reactant M$o\llap{/}$ller state propagation. }
    \label{fig:comb_dynam}
\end{figure*}

The correlation function in this example is defined as $C_{+k,-k}(t) = \bra{^{+}\Psi^{k}_{-}}\exp{-iHt}\ket{^{-}\Psi^{k}_{+}}$. In order to calculate the correlation function at time $t$ we need to propagate the reactant M$o\llap{/}$ller state till time $t$ and take the inner product with the product M$o\llap{/}$ller state. Fig. \ref{fig:comb_dynam}(a) the peak of the wave-packet has progressed towards the interaction region and since the the wavepacket is composed of range of different $k$ values the wavepacket gets broadened (Since, $V(x)=0$ in this region). However, no informative data has been gathered at this stage, as the wave-packet has not yet entered the interaction region. 
The quantum simulation results can be obtained by creating a quantum circuit that only uses the state preparation $\ket{q_{r}}$ quantum registers, and initiate the reactant M$o\llap{/}$ller state, then we apply the propagator and impliment this quantum circuit using the state vector simulator in Qiskit. 

\begin{figure*}[t]
    \centering
    \includegraphics[width=0.9\linewidth]{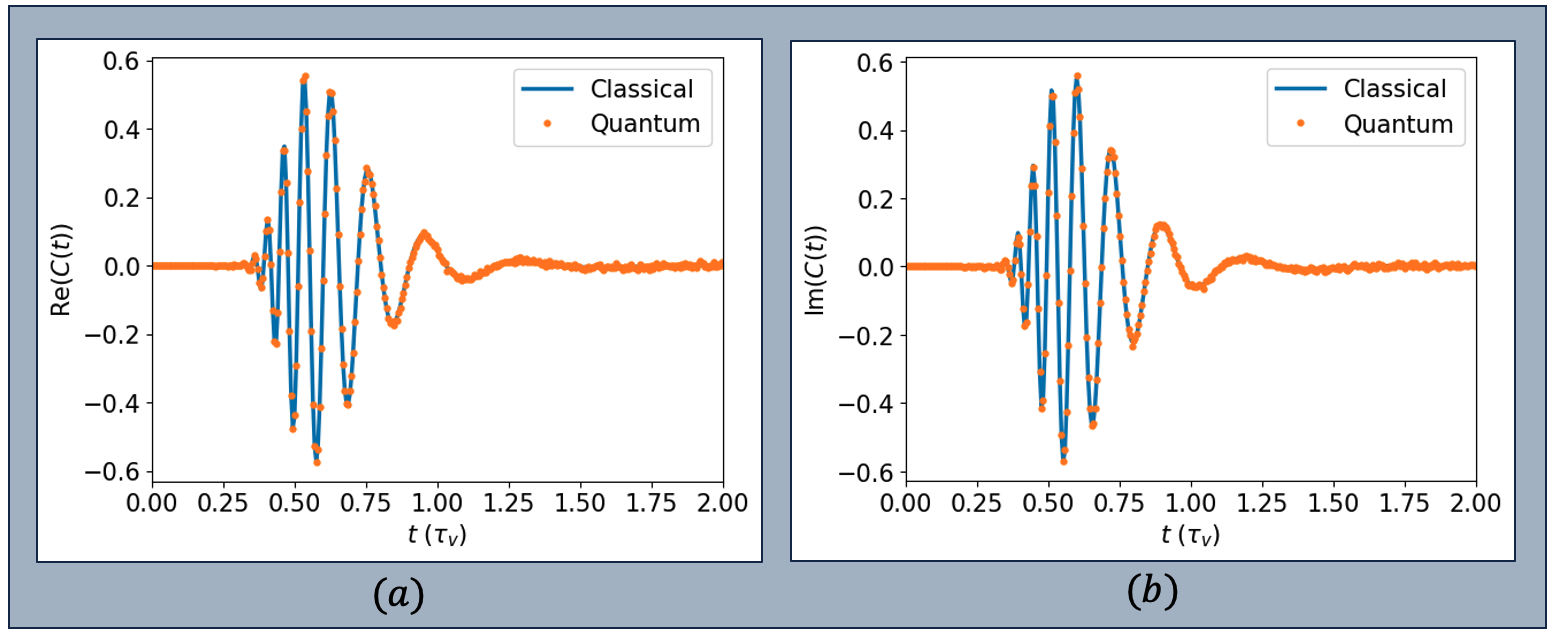}
    \caption{(a) Real and (b) Imaginary contributions to the correlation function at time $t$. The solid blue curve corresponds to the classical simulation results and the orange discrete points is the result from quantum simulation.}
    \label{fig:comb_corr}
\end{figure*}

Fig. \ref{fig:comb_dynam}(b) shows the scaled absolute value of the reactant propagated wavepacket at $t=0.24\tau_{v}$ The wave packet's higher momentum components have now entered the interaction region, engaging with the barrier. In the course of this interaction, the overall energy within the well area rises, characterized by an exchange of kinetic energy for potential energy. No observable signs of bifurcation are evident. It's evident from Fig. \ref{fig:comb_dynam}(c) that at time $t=0.56 \tau_{v}$ the wavepacket post collision begins to exit the interaction region, having acquired information about the potential. Concluding at $t=0.97\tau_{v}$ in Fig. \ref{fig:comb_dynam}(d) the calculation is essentially complete, with only the lower momentum components yet to exit. The correlation function $C_{-k,k}(t)$ is plotted in Fig. \ref{fig:comb_corr} from the figure it's evident that the correlation function was initially zero since the wavepackets were propagating in the opposite directions but once the reactant M$o\llap{/}$ller state get's reflected from the semi-infinite well the correlation function becomes significant and as soon as most of the reflected resultant wavepacket leaves the region around time 1.25 $\tau_{v}$ the correlation function becomes zero and stays the same. 

\begin{figure}[h!]
    \centering
    \includegraphics[width=0.85\linewidth]{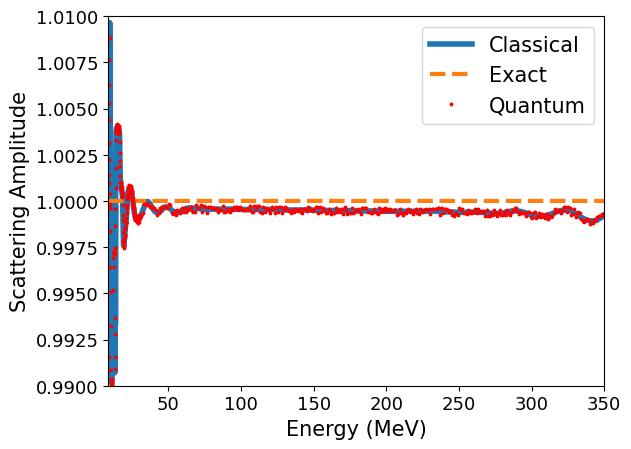}
    \caption{Amplitude of the Scattering matrix element $S_{-k,k}(E)$. Blue curve corresponds to classical numerical calculation, the dashed orange curve denotes the exact expected analytical result and the red circular markers represents result from the quantum simulation.}
    \label{fig:1d_scat_E}
\end{figure} 

Fig. \ref{fig:1d_scat_E} plots the amplitude of the scattering matrix element. Since the entire waveform should be reflected from the potential barrier we anticipate the scattering amplitude should be equal to one. However, the lower sampling rate fails to align with this expectation due to a suboptimal choice of step size. The lower sampling rate, approximately two samples per femtometer, implies that the square well takes on more of a trapezoidal shape rather than a square one.

Addressing this limitation, increasing the sampling rate by higher orders of magnitude rectifies the issue, and the anticipated scattering amplitude of one should be achieved. It's essential to note that both solutions reveal a ringing effect at the upper and lower energy limits, which remains consistent regardless of the sampling rate. 
The occurrence of this ringing phenomenon is attributed to the division by the product of the expansion coefficients $\eta_{\pm}(k(E))$, known for their small values in these energy region. In the next example we include vibrational degrees of freedom and apply the proposed algorithm to calculated scattering matrix elements and reaction probabilities of the co-linear hydrogen exchange reaction.

\subsection{Co-Linear Hydrogen Exchange Reaction}
\label{hydrogen_exchange}
\textbf{Co-Linear Hydrogen Exchange Reaction}

Given it's simplistic nature $H+H_{2}$ chemical reaction is conventionally recognized as the "Hydrogen atom"\cite{schatz1996scattering} of chemical reactions  has held significant importance in theoretical chemistry. In 1929 London, Eyring, and Polanyi demonstrated the existence of a potential energy barrier\cite{wang2000several} to the reaction by providing an approximate solution to the electronic Schrödinger equation. Since then the availability of better computational tools along with precise theoretical models fueled the advancement of the field forward which eventually helped to understand different experimental observations\cite{manolopoulos1993transition}. 
\begin{equation}
    H_{a}+H_{b}H_{c}(v)\rightarrow H_{a}H_{b}(v')+H_{c}
    \label{eq:h3_reaction}
\end{equation}

\begin{figure}[h]
    \centering
    \includegraphics[width=.85\linewidth]{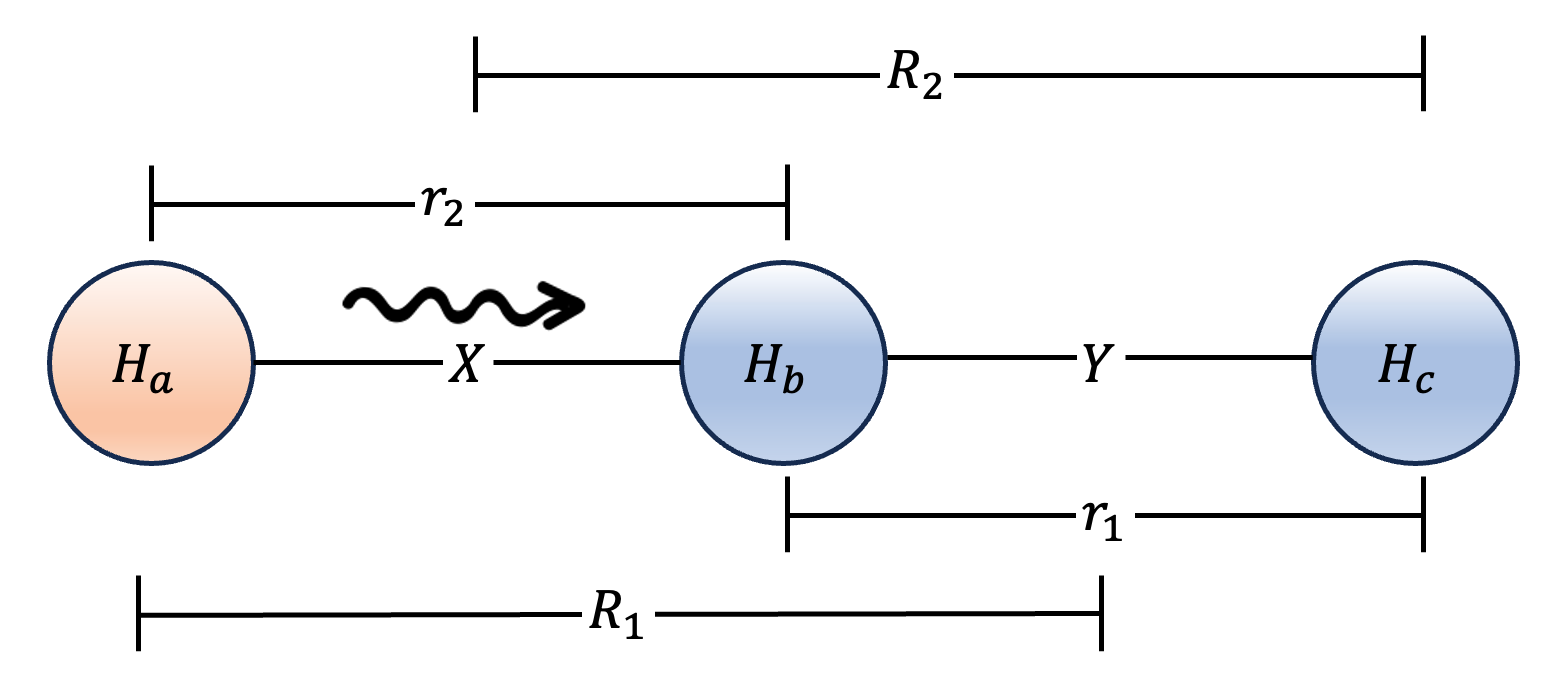}
    \caption{Illustration showing linear arrangement of the three distinguishable  Hydrogen atoms $H_{a},H_{b}$ abd $H_{c}$. Reactant channel $(R_{1},r_{1})$ and product channel $(R_{2},r_{2})$ co-ordinates are along with the bond co-ordinates $(X,Y)$ are shown. The arrow denotes the direction Hydrogen atom $H_{a}$ is travelling.}
    \label{fig:h3_coordinate}
\end{figure}

In this section we employ the proposed quantum algorithm to calculate scattering matrix elements of the co-linear hydrogen exchange reaction Eq. \ref{eq:h3_reaction}. The reaction involves in-elastic scattering between distinguishable hydrogen atom and hydrogen molecule. Hydrogen exchange happens during the chemical reaction where one of the hydrogen atoms substitutes another hydrogen atom. Rotationally averaged scattering matrix components are calculated using the time-dependent technique based on Moller operator formulation of scattering theory (Section \ref{td_theory}). In this chemical process it proves benificial to consider three co-ordinate systems as depicted in Fig. \ref{fig:h3_coordinate}. The co-ordinates $R_{1},r_{1}$ ($R_{2},r_{2}$) is ideally tailored for describing dynamics within the reactant (product) arrangement channel I (II).  In channel I the dynamics is solely governed by the asymptotic hamiltoninan $H^{(1)}_{0}$ Eq. \ref{eq:reac_asymp} (Which is separable into translation between $H_{a}$ and COM of $H_{b}H_{c}$ + Internal vibration in $H_{b}H_{c}$). In a similar fashion coordinates labeled $R_{2}$ and $r_{2}$ in Fig. \ref{fig:h3_coordinate} best describes the dynamics in the product channel $H_{a}H_{b}(v')+H_{c}$ and in the large limit of $R_{2}$ the dynamics is explained by asymptotic hamiltoninan $H^{(2)}_{0}$ Eq. \ref{eq:prod_asymp} ( Which is separable into translation between $H_{c}$ and COM of $H_{a}H_{b}$ + Internal vibration in $H_{a}H_{b}$). 

\begin{equation}
    H^{(1)}_{0}(R_{1},r_{1}) = H^{(1)}_{rel}(R_{1}) +  H^{(1)}_{vib}(r_{1}) 
    \label{eq:reac_asymp}
\end{equation}

\begin{equation}
    H^{(2)}_{0}(R_{2},r_{2}) = H^{(2)}_{rel}(R_{2}) +  H^{(1)}_{vib}(r_{2})
    \label{eq:prod_asymp}
\end{equation}

\begin{figure}[h!]
    \centering
    \includegraphics[width=0.75\linewidth]{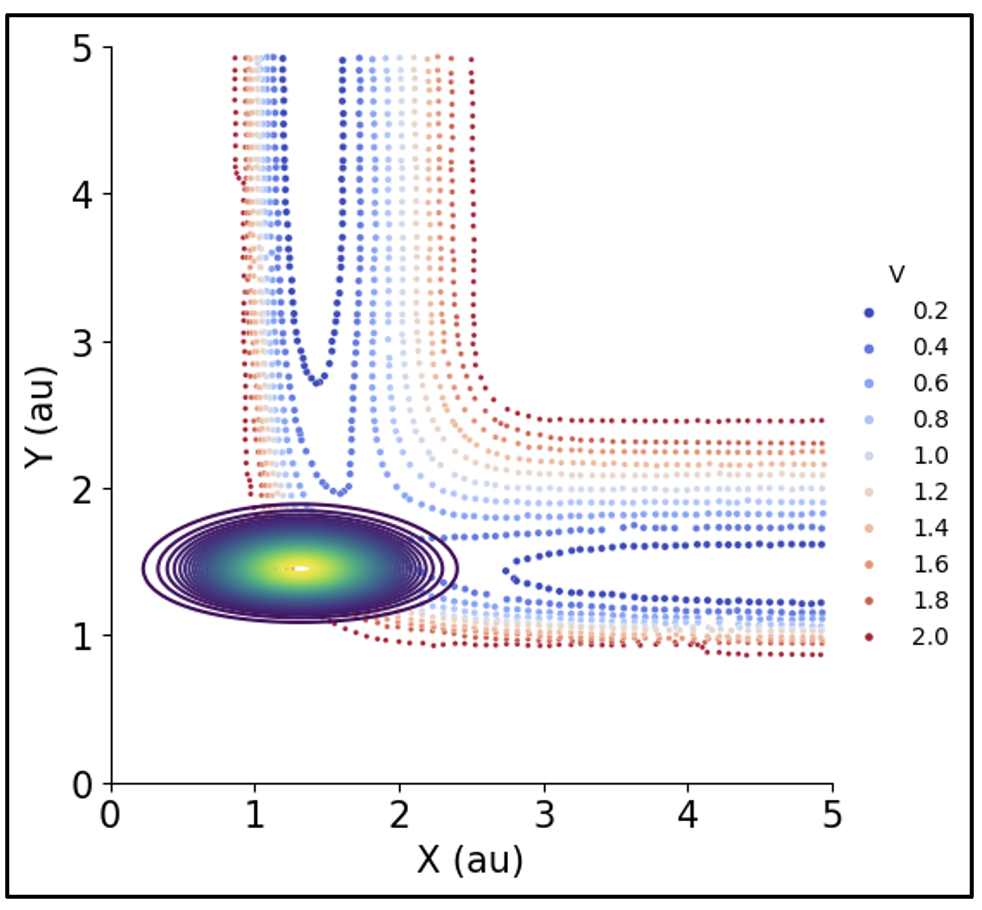}
    \caption{Illustration showing probability density ($|\ket{\psi^{1,0}_{in}(X,Y)}|^{2}$) at time $t=0$  of the reactant channel wavepacket described in Eq. \protect\ref{eq: in_wfn}. The contour plot of the LSTH PES is also shown.}
    \label{fig:h3_init_wfn}
\end{figure}

The third co-ordinate system $X$ and $Y$ as shown in Fig. \ref{fig:h3_coordinate} is used to describe Hamiltonian in the interaction region. Being a co-linear system the transformation is straight forward is given by the relation Eq. \ref{transform}\cite{weeks1994time}. The dynamics in the interaction region is governed by the kinetically coupled hamiltonian\cite{bae1982review} described in Eq. \ref{eq: comb_ham}. $p_{X}$ and $p_{Y}$ represents momentum in the bond co-ordinates and $V(X,Y)$ denotes the Liu-Siegbahn-Truhlar-Horowitz (LSTH)\cite{liu1978LSTH,truhlar1978functional,frishman1997distributed} potential energy surface (PES) as shown in Fig. \ref{fig:h3_init_wfn}. 

\begin{equation}
    \begin{pmatrix}
        R_{1}-\frac{1}{2} r_{1} \\
        r_{1}
    \end{pmatrix} =     \begin{pmatrix}
        X \\
        Y
    \end{pmatrix} =     \begin{pmatrix}
        r_{2} \\
        R_{2}-\frac{1}{2} r_{2}
    \end{pmatrix}
    \label{transform}
\end{equation}

\begin{equation}
    H = \frac{p_{X}^{2}}{m_{H}} - \frac{p_{X}p_{Y}}{m_{H}} + \frac{p_{Y}^{2}}{m_{H}} + V(X,Y)
    \label{eq: comb_ham}
\end{equation}

The reactant wavepacket is created by calculating the direct product between an arbitrary linear combination of one-dimensional plane waves characterizing the relative motion of the atom $H_{a}$ and the diatom $H_{b}H_{c}(v)$, and a single vibrational eigenstate, (v) of the diatom $H_{b}H_{c}$. Similarly an outgoing channel wavepacket that corresponds to $H_{a}H_{b}(v')+H_{c}$ is calculated as a direct product between an arbitrary linear combination of one-dimensional plane waves characterizing the relative motion of the atom $H_{c}$ and the diatom $H_{a}H_{b}(v')$ and a single vibrational eigenstate, $(v')$ of the diatom $H_{a}H_{b}$. The reaction channel wavepacket $\ket{\psi^{1,0}_{in}}$ at time $t=0$  is created by directly multiplying the translational and vibrational wavefunctions. In Eq. \ref{eq: in_wfn} $N$ corresponds to the normalization constant, the superscript in $\psi^{\gamma,v}$ denotes the channel $\gamma$ and vibrational quantum number of $v$, A slice of LSTH- PES taken at a large value of $R_{1}$ defines the asymptotic diatomic potential. $H^{1}_{vib}(r_{1})$ is expressed in the Morse eigenbasis and diagonalized to extract the vibrational wavefunction $(\ket{\psi^{1,0}_{vib}(r_{1})})$. The product channel wavepacket can be defined in a similar fashion (Please refer to the Supplementary Information). 

\begin{equation}
    \ket{\psi^{1,0}_{in}(R_{1},r_{1})}  = N \ket{\psi^{1,0}_{vib}(r_{1})} e^{-\Delta k^{2}_{1}(R_{1}-R^{0}_{1})^{2}+ik^{0}_{1}(R_{1}-R^{0}_{1}))} 
    \label{eq: in_wfn}
\end{equation}

\begin{figure*}[t]
    \centering
    \includegraphics[width=0.85\linewidth]{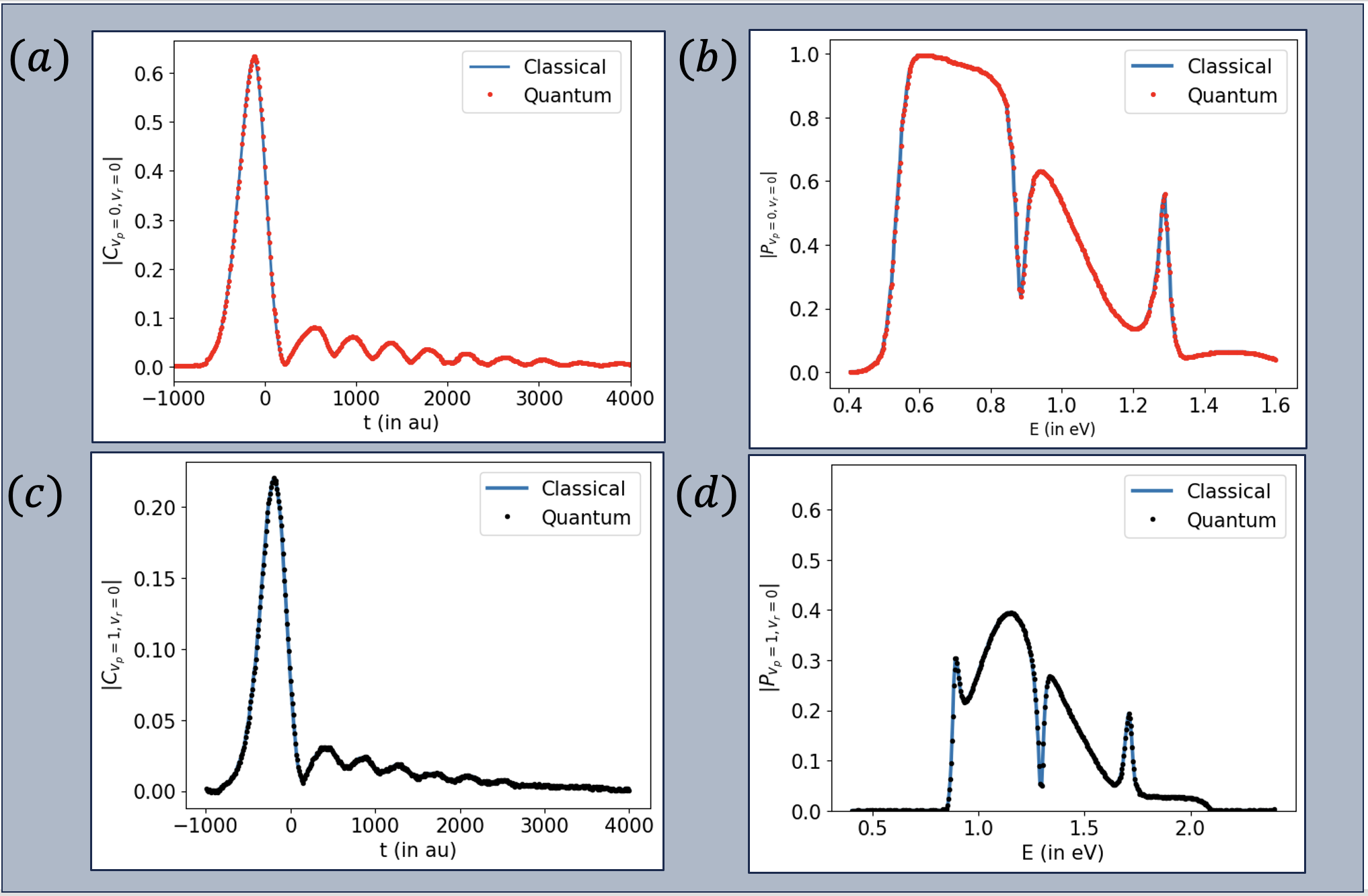}
    \caption{(a) Absolute value of the correlation function $C_{v_{p}=0,v_{r}=0}(t)$,(c) Absolute value of the correlation function $C_{v_{p}=1,v_{r}=0}(t)$.(b) The probability of the reaction $P_{v_{p}=0,v_{r}=0}$. (d) The probability of the reaction $P_{v_{p}=1,v_{r}=0}$. The solid blue curve corresponds to the classical simulation results and the red (black) discrete points is the result from quantum simulation.}
    \label{fig:comb_h3}
\end{figure*}

 We calculate S-matrix elements for the two inelastic exchange reactions, where we start with $v=0$ in the reactant wavepacket and $v'={0,1}$ in the product wavepackets. In this case the number of qubits required to represent the quantum state in algorithm 10  (2 quibts corresponds to vibrational encoding and the rest to represent $2^{8}$ $\eta_{\pm}(k_{\gamma]})$. The quantum algorithm is executed at each time $t$ and 
the correlation function is calculated between the reactant $(\ket{\psi^{\gamma}_{+}})$ and product $(\ket{\psi^{\gamma'}_{-}})$ M$o\llap{/}$ller states. Fig. \ref{fig:comb_h3}(a) (Fig. \ref{fig:comb_h3}(c)) shows the correlation function for $H_{a}+H_{b}H_{c}(v=0)\rightarrow H_{a}H_{b}(v'=0(1))+H_{c}$ case. It is evident that the quantum simulation results goes well with the exact classical simulation. The fourier transform of the calculated correlation function is used to calculate scattering matrix elements. Fig. \ref{fig:comb_h3}(b) and Fig. \ref{fig:comb_h3}(d) plots the transmission coefficient or the reaction probability which is defied as absolute value square of the corresponding scattering matrix elements $P_{v',v}(E) = \abs{S^{v',v}_{+k,-k}}^{2}$. These computations offer an authentic portrayal of the reaction dynamics, capturing all the important aspects in the reaction probabilities. As established in prior research\cite{schatz1973role}, the interplay between direct and resonant components plays a pivotal role in shaping the overall curves depicted in Fig. \ref{fig:comb_h3}(b). The abrupt fluctuations around 0.6 and 1.0 eV arise due to the existence of internal excitation resonances at these energy levels.

\section*{Conclusion}

In this study, we introduced a quantum algorithm based on the TD M$o\llap{/}$ller to compute scattering matrix elements for both elastic and inelastic scattering processes. To address the requirements for calculating the correlation function, we developed a Modified Hadamard test. The proposed algorithm was successfully applied to two distinct physical problems. Initially, we tackled the 1D semi-infinite well, serving as an approximation for the AV18 $^{1}S_{0}$ 2-nuclei scattering potential. Subsequently, we extended the application to the co-linear Hydrogen exchange reaction. Remarkably, the quantum simulation results demonstrated excellent agreement with classical results in both cases.

Moreover, the algorithm presents two significant sources of errors deserving careful consideration. The first originates from sampling errors associated with the Hadamard test. This error is tied to the estimation of the expected value of the correlation function based on a finite number of sample measurements. Through the application of Chernoff bound,\cite{chernoff1981note} we established that the required number of samples to estimate the expected value with an absolute error $\epsilon$ scales as $\mathcal{O}(\frac{1}{\epsilon^{2}})$.

The other source of error comes from higher order Trotter approximation of the propagator ($\exp(-iHt)$). Accurately assessing Trotter approximation errors is crucial for optimizing Hamiltonian simulations. Efforts\cite{childs2021theory} have been made to devise a theory that leverages the commutativity of operator summands to yield more tighter error bounds. Despite these endeavors, challenges persist due to the intricacies involved in obtaining analytical expressions for the error bounds and understanding the impact of the approximation on quantum simulation errors. In the future we would like to extend this formalism from the co-linear Hydrogen exchange to include the rotational aspect in the scattering process. Thus would open up new avenues to deeply understand molecular processes which exibits quantum coherent control.

\begin{acknowledgement}

The authors would like to thank Dr. Manas Sajjan for insightful discussions. We would like also to acknowledge funding from the Office
of Science through the Quantum Science Center (QSC), a
National Quantum Information Science Research Center, the
U.S. Department of Energy (DOE) (Office of Basic Energy
Sciences), under Award No. DE-SC0019215, and the National
Science Foundation under Award No. 1955907.

\end{acknowledgement}


\section{Supporting Information}
 \textbf{Supporting Information:} The supplementary information provides a detailed expressions for the scattering matrix elements, a thorough look at the quantum algorithm, explanations about Hamiltonian encoding, quantum dynamics parameters, and error analysis.

\bibliography{achemso-demo}

\providecommand{\noopsort}[1]{}\providecommand{\singleletter}[1]{#1}%
\providecommand{\latin}[1]{#1}
\makeatletter
\providecommand{\doi}
  {\begingroup\let\do\@makeother\dospecials
  \catcode`\{=1 \catcode`\}=2 \doi@aux}
\providecommand{\doi@aux}[1]{\endgroup\texttt{#1}}
\makeatother
\providecommand*\mcitethebibliography{\thebibliography}
\csname @ifundefined\endcsname{endmcitethebibliography}  {\let\endmcitethebibliography\endthebibliography}{}
\begin{mcitethebibliography}{54}
\providecommand*\natexlab[1]{#1}
\providecommand*\mciteSetBstSublistMode[1]{}
\providecommand*\mciteSetBstMaxWidthForm[2]{}
\providecommand*\mciteBstWouldAddEndPuncttrue
  {\def\EndOfBibitem{\unskip.}}
\providecommand*\mciteBstWouldAddEndPunctfalse
  {\let\EndOfBibitem\relax}
\providecommand*\mciteSetBstMidEndSepPunct[3]{}
\providecommand*\mciteSetBstSublistLabelBeginEnd[3]{}
\providecommand*\EndOfBibitem{}
\mciteSetBstSublistMode{f}
\mciteSetBstMaxWidthForm{subitem}{(\alph{mcitesubitemcount})}
\mciteSetBstSublistLabelBeginEnd
  {\mcitemaxwidthsubitemform\space}
  {\relax}
  {\relax}

\bibitem[Arute \latin{et~al.}(2019)Arute, Arya, Babbush, Bacon, Bardin, Barends, Biswas, Boixo, Brandao, Buell, \latin{et~al.} others]{arute2019quantum}
Arute,~F.; Arya,~K.; Babbush,~R.; Bacon,~D.; Bardin,~J.~C.; Barends,~R.; Biswas,~R.; Boixo,~S.; Brandao,~F.~G.; Buell,~D.~A.; others Quantum supremacy using a programmable superconducting processor. \emph{Nature} \textbf{2019}, \emph{574}, 505--510\relax
\mciteBstWouldAddEndPuncttrue
\mciteSetBstMidEndSepPunct{\mcitedefaultmidpunct}
{\mcitedefaultendpunct}{\mcitedefaultseppunct}\relax
\EndOfBibitem
\bibitem[Kim \latin{et~al.}(2023)Kim, Eddins, Anand, Wei, Van Den~Berg, Rosenblatt, Nayfeh, Wu, Zaletel, Temme, \latin{et~al.} others]{kim2023evidence}
Kim,~Y.; Eddins,~A.; Anand,~S.; Wei,~K.~X.; Van Den~Berg,~E.; Rosenblatt,~S.; Nayfeh,~H.; Wu,~Y.; Zaletel,~M.; Temme,~K.; others Evidence for the utility of quantum computing before fault tolerance. \emph{Nature} \textbf{2023}, \emph{618}, 500--505\relax
\mciteBstWouldAddEndPuncttrue
\mciteSetBstMidEndSepPunct{\mcitedefaultmidpunct}
{\mcitedefaultendpunct}{\mcitedefaultseppunct}\relax
\EndOfBibitem
\bibitem[Bharti \latin{et~al.}(2022)Bharti, Cervera-Lierta, Kyaw, Haug, Alperin-Lea, Anand, Degroote, Heimonen, Kottmann, Menke, \latin{et~al.} others]{bharti2022noisy}
Bharti,~K.; Cervera-Lierta,~A.; Kyaw,~T.~H.; Haug,~T.; Alperin-Lea,~S.; Anand,~A.; Degroote,~M.; Heimonen,~H.; Kottmann,~J.~S.; Menke,~T.; others Noisy intermediate-scale quantum algorithms. \emph{Reviews of Modern Physics} \textbf{2022}, \emph{94}, 015004\relax
\mciteBstWouldAddEndPuncttrue
\mciteSetBstMidEndSepPunct{\mcitedefaultmidpunct}
{\mcitedefaultendpunct}{\mcitedefaultseppunct}\relax
\EndOfBibitem
\bibitem[Sajjan \latin{et~al.}(2022)Sajjan, Li, Selvarajan, Sureshbabu, Kale, Gupta, Singh, and Kais]{sajjan2022quantum}
Sajjan,~M.; Li,~J.; Selvarajan,~R.; Sureshbabu,~S.~H.; Kale,~S.~S.; Gupta,~R.; Singh,~V.; Kais,~S. Quantum machine learning for chemistry and physics. \emph{Chemical Society Reviews} \textbf{2022}, \relax
\mciteBstWouldAddEndPunctfalse
\mciteSetBstMidEndSepPunct{\mcitedefaultmidpunct}
{}{\mcitedefaultseppunct}\relax
\EndOfBibitem
\bibitem[Sawaya \latin{et~al.}(2021)Sawaya, Paesani, and Tabor]{sawaya2021near}
Sawaya,~N.~P.; Paesani,~F.; Tabor,~D.~P. Near-and long-term quantum algorithmic approaches for vibrational spectroscopy. \emph{Physical Review A} \textbf{2021}, \emph{104}, 062419\relax
\mciteBstWouldAddEndPuncttrue
\mciteSetBstMidEndSepPunct{\mcitedefaultmidpunct}
{\mcitedefaultendpunct}{\mcitedefaultseppunct}\relax
\EndOfBibitem
\bibitem[Bruschi \latin{et~al.}(2024)Bruschi, Gallina, and Fresch]{bruschi2024quantum}
Bruschi,~M.; Gallina,~F.; Fresch,~B. A Quantum Algorithm from Response Theory: Digital Quantum Simulation of Two-Dimensional Electronic Spectroscopy. \emph{The Journal of Physical Chemistry Letters} \textbf{2024}, \emph{15}, 1484--1492\relax
\mciteBstWouldAddEndPuncttrue
\mciteSetBstMidEndSepPunct{\mcitedefaultmidpunct}
{\mcitedefaultendpunct}{\mcitedefaultseppunct}\relax
\EndOfBibitem
\bibitem[Johri \latin{et~al.}(2017)Johri, Steiger, and Troyer]{johri2017entanglement}
Johri,~S.; Steiger,~D.~S.; Troyer,~M. Entanglement spectroscopy on a quantum computer. \emph{Physical Review B} \textbf{2017}, \emph{96}, 195136\relax
\mciteBstWouldAddEndPuncttrue
\mciteSetBstMidEndSepPunct{\mcitedefaultmidpunct}
{\mcitedefaultendpunct}{\mcitedefaultseppunct}\relax
\EndOfBibitem
\bibitem[Xia \latin{et~al.}(2017)Xia, Bian, and Kais]{xia2017electronic}
Xia,~R.; Bian,~T.; Kais,~S. Electronic structure calculations and the Ising Hamiltonian. \emph{The Journal of Physical Chemistry B} \textbf{2017}, \emph{122}, 3384--3395\relax
\mciteBstWouldAddEndPuncttrue
\mciteSetBstMidEndSepPunct{\mcitedefaultmidpunct}
{\mcitedefaultendpunct}{\mcitedefaultseppunct}\relax
\EndOfBibitem
\bibitem[Xia and Kais(2018)Xia, and Kais]{xia2018quantum}
Xia,~R.; Kais,~S. Quantum machine learning for electronic structure calculations. \emph{Nature communications} \textbf{2018}, \emph{9}, 4195\relax
\mciteBstWouldAddEndPuncttrue
\mciteSetBstMidEndSepPunct{\mcitedefaultmidpunct}
{\mcitedefaultendpunct}{\mcitedefaultseppunct}\relax
\EndOfBibitem
\bibitem[Sajjan \latin{et~al.}(2023)Sajjan, Gupta, Kale, Singh, Kumaran, and Kais]{sajjan2023physics}
Sajjan,~M.; Gupta,~R.; Kale,~S.~S.; Singh,~V.; Kumaran,~K.; Kais,~S. Physics-Inspired Quantum Simulation of Resonating Valence Bond States A Prototypical Template for a Spin-Liquid Ground State. \emph{The Journal of Physical Chemistry A} \textbf{2023}, \emph{127}, 8751--8764\relax
\mciteBstWouldAddEndPuncttrue
\mciteSetBstMidEndSepPunct{\mcitedefaultmidpunct}
{\mcitedefaultendpunct}{\mcitedefaultseppunct}\relax
\EndOfBibitem
\bibitem[Ollitrault \latin{et~al.}(2020)Ollitrault, Baiardi, Reiher, and Tavernelli]{ollitrault2020hardware}
Ollitrault,~P.~J.; Baiardi,~A.; Reiher,~M.; Tavernelli,~I. Hardware efficient quantum algorithms for vibrational structure calculations. \emph{Chemical science} \textbf{2020}, \emph{11}, 6842--6855\relax
\mciteBstWouldAddEndPuncttrue
\mciteSetBstMidEndSepPunct{\mcitedefaultmidpunct}
{\mcitedefaultendpunct}{\mcitedefaultseppunct}\relax
\EndOfBibitem
\bibitem[Bravyi \latin{et~al.}(2019)Bravyi, Gosset, K{\"o}nig, and Temme]{bravyi2019approximation}
Bravyi,~S.; Gosset,~D.; K{\"o}nig,~R.; Temme,~K. Approximation algorithms for quantum many-body problems. \emph{Journal of Mathematical Physics} \textbf{2019}, \emph{60}\relax
\mciteBstWouldAddEndPuncttrue
\mciteSetBstMidEndSepPunct{\mcitedefaultmidpunct}
{\mcitedefaultendpunct}{\mcitedefaultseppunct}\relax
\EndOfBibitem
\bibitem[Hu \latin{et~al.}(2020)Hu, Xia, and Kais]{hu2020quantum}
Hu,~Z.; Xia,~R.; Kais,~S. A quantum algorithm for evolving open quantum dynamics on quantum computing devices. \emph{Scientific reports} \textbf{2020}, \emph{10}, 3301\relax
\mciteBstWouldAddEndPuncttrue
\mciteSetBstMidEndSepPunct{\mcitedefaultmidpunct}
{\mcitedefaultendpunct}{\mcitedefaultseppunct}\relax
\EndOfBibitem
\bibitem[Althorpe and Clary(2003)Althorpe, and Clary]{althorpe2003quantum}
Althorpe,~S.~C.; Clary,~D.~C. Quantum scattering calculations on chemical reactions. \emph{Annual review of physical chemistry} \textbf{2003}, \emph{54}, 493--529\relax
\mciteBstWouldAddEndPuncttrue
\mciteSetBstMidEndSepPunct{\mcitedefaultmidpunct}
{\mcitedefaultendpunct}{\mcitedefaultseppunct}\relax
\EndOfBibitem
\bibitem[Fu \latin{et~al.}(2017)Fu, Shan, Zhang, and Clary]{fu2017recent}
Fu,~B.; Shan,~X.; Zhang,~D.~H.; Clary,~D.~C. Recent advances in quantum scattering calculations on polyatomic bimolecular reactions. \emph{Chemical Society Reviews} \textbf{2017}, \emph{46}, 7625--7649\relax
\mciteBstWouldAddEndPuncttrue
\mciteSetBstMidEndSepPunct{\mcitedefaultmidpunct}
{\mcitedefaultendpunct}{\mcitedefaultseppunct}\relax
\EndOfBibitem
\bibitem[Guitou \latin{et~al.}(2015)Guitou, Spielfiedel, Rodionov, Yakovleva, Belyaev, Merle, Thevenin, and Feautrier]{guitou2015quantum}
Guitou,~M.; Spielfiedel,~A.; Rodionov,~D.; Yakovleva,~S.; Belyaev,~A.; Merle,~T.; Thevenin,~F.; Feautrier,~N. Quantum chemistry and nuclear dynamics as diagnostic tools for stellar atmosphere modeling. \emph{Chemical Physics} \textbf{2015}, \emph{462}, 94--103\relax
\mciteBstWouldAddEndPuncttrue
\mciteSetBstMidEndSepPunct{\mcitedefaultmidpunct}
{\mcitedefaultendpunct}{\mcitedefaultseppunct}\relax
\EndOfBibitem
\bibitem[Madronich and Flocke(1999)Madronich, and Flocke]{madronich1999role}
Madronich,~S.; Flocke,~S. The role of solar radiation in atmospheric chemistry. \emph{Environmental photochemistry} \textbf{1999}, 1--26\relax
\mciteBstWouldAddEndPuncttrue
\mciteSetBstMidEndSepPunct{\mcitedefaultmidpunct}
{\mcitedefaultendpunct}{\mcitedefaultseppunct}\relax
\EndOfBibitem
\bibitem[Levine(1969)]{Levine}
Levine,~R. \emph{Quantum mecahnics of molecular rate processes}; Oxford Univ. Press , Oxford, 1969\relax
\mciteBstWouldAddEndPuncttrue
\mciteSetBstMidEndSepPunct{\mcitedefaultmidpunct}
{\mcitedefaultendpunct}{\mcitedefaultseppunct}\relax
\EndOfBibitem
\bibitem[Zhang and Guo(2016)Zhang, and Guo]{zhang2016recent}
Zhang,~D.~H.; Guo,~H. Recent advances in quantum dynamics of bimolecular reactions. \emph{Annual review of physical chemistry} \textbf{2016}, \emph{67}, 135--158\relax
\mciteBstWouldAddEndPuncttrue
\mciteSetBstMidEndSepPunct{\mcitedefaultmidpunct}
{\mcitedefaultendpunct}{\mcitedefaultseppunct}\relax
\EndOfBibitem
\bibitem[Pozdneev(2019)]{pozdneev2019application}
Pozdneev,~S.~A. Application of quantum scattering theory in calculation of the simplest chemical reactions (dissociative attachment, dissociation, and recombination). \emph{Technical Physics} \textbf{2019}, \emph{64}, 749--756\relax
\mciteBstWouldAddEndPuncttrue
\mciteSetBstMidEndSepPunct{\mcitedefaultmidpunct}
{\mcitedefaultendpunct}{\mcitedefaultseppunct}\relax
\EndOfBibitem
\bibitem[Krems(2008)]{krems2008cold}
Krems,~R.~V. Cold controlled chemistry. \emph{Physical Chemistry Chemical Physics} \textbf{2008}, \emph{10}, 4079--4092\relax
\mciteBstWouldAddEndPuncttrue
\mciteSetBstMidEndSepPunct{\mcitedefaultmidpunct}
{\mcitedefaultendpunct}{\mcitedefaultseppunct}\relax
\EndOfBibitem
\bibitem[Mellish \latin{et~al.}(2007)Mellish, Kj{\ae}rgaard, Julienne, and Wilson]{mellish2007quantum}
Mellish,~A.~S.; Kj{\ae}rgaard,~N.; Julienne,~P.~S.; Wilson,~A.~C. Quantum scattering of distinguishable bosons using an ultracold-atom collider. \emph{Physical Review A} \textbf{2007}, \emph{75}, 020701\relax
\mciteBstWouldAddEndPuncttrue
\mciteSetBstMidEndSepPunct{\mcitedefaultmidpunct}
{\mcitedefaultendpunct}{\mcitedefaultseppunct}\relax
\EndOfBibitem
\bibitem[Shapiro and Brumer(1997)Shapiro, and Brumer]{shapiro1997quantum}
Shapiro,~M.; Brumer,~P. Quantum control of chemical reactions. \emph{Journal of the Chemical Society, Faraday Transactions} \textbf{1997}, \emph{93}, 1263--1277\relax
\mciteBstWouldAddEndPuncttrue
\mciteSetBstMidEndSepPunct{\mcitedefaultmidpunct}
{\mcitedefaultendpunct}{\mcitedefaultseppunct}\relax
\EndOfBibitem
\bibitem[Kale \latin{et~al.}(2021)Kale, Chen, and Kais]{kale2021constructive}
Kale,~S.~S.; Chen,~Y.~P.; Kais,~S. Constructive Quantum Interference in Photochemical Reactions. \emph{Journal of Chemical Theory and Computation} \textbf{2021}, \emph{17}, 7822--7826\relax
\mciteBstWouldAddEndPuncttrue
\mciteSetBstMidEndSepPunct{\mcitedefaultmidpunct}
{\mcitedefaultendpunct}{\mcitedefaultseppunct}\relax
\EndOfBibitem
\bibitem[Aoiz and Zare(2018)Aoiz, and Zare]{aoiz2018quantum}
Aoiz,~F.~J.; Zare,~R.~N. Quantum interference in chemical reactions. \emph{Physics Today} \textbf{2018}, \emph{71}, 70--71\relax
\mciteBstWouldAddEndPuncttrue
\mciteSetBstMidEndSepPunct{\mcitedefaultmidpunct}
{\mcitedefaultendpunct}{\mcitedefaultseppunct}\relax
\EndOfBibitem
\bibitem[Bian and Kais(2021)Bian, and Kais]{bian2021quantum}
Bian,~T.; Kais,~S. Quantum computing for atomic and molecular resonances. \emph{The Journal of Chemical Physics} \textbf{2021}, \emph{154}\relax
\mciteBstWouldAddEndPuncttrue
\mciteSetBstMidEndSepPunct{\mcitedefaultmidpunct}
{\mcitedefaultendpunct}{\mcitedefaultseppunct}\relax
\EndOfBibitem
\bibitem[Kassal \latin{et~al.}(2008)Kassal, Jordan, Love, Mohseni, and Aspuru-Guzik]{kassal2008polynomial}
Kassal,~I.; Jordan,~S.~P.; Love,~P.~J.; Mohseni,~M.; Aspuru-Guzik,~A. Polynomial-time quantum algorithm for the simulation of chemical dynamics. \emph{Proceedings of the National Academy of Sciences} \textbf{2008}, \emph{105}, 18681--18686\relax
\mciteBstWouldAddEndPuncttrue
\mciteSetBstMidEndSepPunct{\mcitedefaultmidpunct}
{\mcitedefaultendpunct}{\mcitedefaultseppunct}\relax
\EndOfBibitem
\bibitem[Du \latin{et~al.}(2021)Du, Vary, Zhao, and Zuo]{du2021quantum}
Du,~W.; Vary,~J.~P.; Zhao,~X.; Zuo,~W. Quantum simulation of nuclear inelastic scattering. \emph{Physical Review A} \textbf{2021}, \emph{104}, 012611\relax
\mciteBstWouldAddEndPuncttrue
\mciteSetBstMidEndSepPunct{\mcitedefaultmidpunct}
{\mcitedefaultendpunct}{\mcitedefaultseppunct}\relax
\EndOfBibitem
\bibitem[Xing \latin{et~al.}(2023)Xing, Gomez~Cadavid, Izmaylov, and Tscherbul]{xing2023hybrid}
Xing,~X.; Gomez~Cadavid,~A.; Izmaylov,~A.~F.; Tscherbul,~T.~V. A hybrid quantum-classical algorithm for multichannel quantum scattering of atoms and molecules. \emph{The Journal of Physical Chemistry Letters} \textbf{2023}, \emph{14}, 6224--6233\relax
\mciteBstWouldAddEndPuncttrue
\mciteSetBstMidEndSepPunct{\mcitedefaultmidpunct}
{\mcitedefaultendpunct}{\mcitedefaultseppunct}\relax
\EndOfBibitem
\bibitem[Bravo-Prieto \latin{et~al.}(2023)Bravo-Prieto, LaRose, Cerezo, Subasi, Cincio, and Coles]{bravo2023variational}
Bravo-Prieto,~C.; LaRose,~R.; Cerezo,~M.; Subasi,~Y.; Cincio,~L.; Coles,~P.~J. Variational quantum linear solver. \emph{Quantum} \textbf{2023}, \emph{7}, 1188\relax
\mciteBstWouldAddEndPuncttrue
\mciteSetBstMidEndSepPunct{\mcitedefaultmidpunct}
{\mcitedefaultendpunct}{\mcitedefaultseppunct}\relax
\EndOfBibitem
\bibitem[Das and Tannor(1990)Das, and Tannor]{das1990time}
Das,~S.; Tannor,~D.~J. Time dependent quantum mechanics using picosecond time steps: Application to predissociation of HeI2. \emph{The Journal of chemical physics} \textbf{1990}, \emph{92}, 3403--3409\relax
\mciteBstWouldAddEndPuncttrue
\mciteSetBstMidEndSepPunct{\mcitedefaultmidpunct}
{\mcitedefaultendpunct}{\mcitedefaultseppunct}\relax
\EndOfBibitem
\bibitem[Weeks and Tannor(1993)Weeks, and Tannor]{weeks1993time}
Weeks,~D.~E.; Tannor,~D.~J. A time-dependent formulation of the scattering matrix using M{\o}ller operators. \emph{Chemical physics letters} \textbf{1993}, \emph{207}, 301--308\relax
\mciteBstWouldAddEndPuncttrue
\mciteSetBstMidEndSepPunct{\mcitedefaultmidpunct}
{\mcitedefaultendpunct}{\mcitedefaultseppunct}\relax
\EndOfBibitem
\bibitem[Tannor and Weeks(1993)Tannor, and Weeks]{tannor1993wave}
Tannor,~D.~J.; Weeks,~D.~E. Wave packet correlation function formulation of scattering theory: The quantum analog of classical S-matrix theory. \emph{The Journal of chemical physics} \textbf{1993}, \emph{98}, 3884--3893\relax
\mciteBstWouldAddEndPuncttrue
\mciteSetBstMidEndSepPunct{\mcitedefaultmidpunct}
{\mcitedefaultendpunct}{\mcitedefaultseppunct}\relax
\EndOfBibitem
\bibitem[Kosloff(1988)]{kosloff1988time}
Kosloff,~R. Time-dependent quantum-mechanical methods for molecular dynamics. \emph{The Journal of Physical Chemistry} \textbf{1988}, \emph{92}, 2087--2100\relax
\mciteBstWouldAddEndPuncttrue
\mciteSetBstMidEndSepPunct{\mcitedefaultmidpunct}
{\mcitedefaultendpunct}{\mcitedefaultseppunct}\relax
\EndOfBibitem
\bibitem[Agmon and Kosloff(1987)Agmon, and Kosloff]{agmon1987dynamics}
Agmon,~N.; Kosloff,~R. Dynamics of two-dimensional diffusional barrier crossing. \emph{Journal of Physical Chemistry} \textbf{1987}, \emph{91}, 1988--1996\relax
\mciteBstWouldAddEndPuncttrue
\mciteSetBstMidEndSepPunct{\mcitedefaultmidpunct}
{\mcitedefaultendpunct}{\mcitedefaultseppunct}\relax
\EndOfBibitem
\bibitem[Zhang(1990)]{zhang1990new}
Zhang,~J.~Z. New method in time-dependent quantum scattering theory: Integrating the wave function in the interaction picture. \emph{The Journal of chemical physics} \textbf{1990}, \emph{92}, 324--331\relax
\mciteBstWouldAddEndPuncttrue
\mciteSetBstMidEndSepPunct{\mcitedefaultmidpunct}
{\mcitedefaultendpunct}{\mcitedefaultseppunct}\relax
\EndOfBibitem
\bibitem[Engel and Metiu(1988)Engel, and Metiu]{engel1988relative}
Engel,~V.; Metiu,~H. The relative kinetic energy distribution of the hydrogen atoms formed by the dissociation of the electronically excited H2 molecule. \emph{The Journal of chemical physics} \textbf{1988}, \emph{89}, 1986--1993\relax
\mciteBstWouldAddEndPuncttrue
\mciteSetBstMidEndSepPunct{\mcitedefaultmidpunct}
{\mcitedefaultendpunct}{\mcitedefaultseppunct}\relax
\EndOfBibitem
\bibitem[Sawaya \latin{et~al.}(2020)Sawaya, Menke, Kyaw, Johri, Aspuru-Guzik, and Guerreschi]{sawaya2020resource}
Sawaya,~N.~P.; Menke,~T.; Kyaw,~T.~H.; Johri,~S.; Aspuru-Guzik,~A.; Guerreschi,~G.~G. Resource-efficient digital quantum simulation of d-level systems for photonic, vibrational, and spin-s Hamiltonians. \emph{npj Quantum Information} \textbf{2020}, \emph{6}, 49\relax
\mciteBstWouldAddEndPuncttrue
\mciteSetBstMidEndSepPunct{\mcitedefaultmidpunct}
{\mcitedefaultendpunct}{\mcitedefaultseppunct}\relax
\EndOfBibitem
\bibitem[Hatano and Suzuki(2005)Hatano, and Suzuki]{hatano2005finding}
Hatano,~N.; Suzuki,~M. \emph{Quantum annealing and other optimization methods}; Springer, 2005; pp 37--68\relax
\mciteBstWouldAddEndPuncttrue
\mciteSetBstMidEndSepPunct{\mcitedefaultmidpunct}
{\mcitedefaultendpunct}{\mcitedefaultseppunct}\relax
\EndOfBibitem
\bibitem[McKay \latin{et~al.}(2018)McKay, Alexander, Bello, Biercuk, Bishop, Chen, Chow, C{\'o}rcoles, Egger, Filipp, \latin{et~al.} others]{mckay2018qiskit}
McKay,~D.~C.; Alexander,~T.; Bello,~L.; Biercuk,~M.~J.; Bishop,~L.; Chen,~J.; Chow,~J.~M.; C{\'o}rcoles,~A.~D.; Egger,~D.; Filipp,~S.; others Qiskit backend specifications for openqasm and openpulse experiments. \emph{arXiv preprint arXiv:1809.03452} \textbf{2018}, \relax
\mciteBstWouldAddEndPunctfalse
\mciteSetBstMidEndSepPunct{\mcitedefaultmidpunct}
{}{\mcitedefaultseppunct}\relax
\EndOfBibitem
\bibitem[Davis(2010)]{davis2010time}
Davis,~B.~S. \emph{Time dependent channel packet calculation of two nucleon scattering matrix elements}; Air Force Institute of Technology, 2010\relax
\mciteBstWouldAddEndPuncttrue
\mciteSetBstMidEndSepPunct{\mcitedefaultmidpunct}
{\mcitedefaultendpunct}{\mcitedefaultseppunct}\relax
\EndOfBibitem
\bibitem[Wiringa \latin{et~al.}(1995)Wiringa, Stoks, and Schiavilla]{wiringa1995accurate}
Wiringa,~R.~B.; Stoks,~V.; Schiavilla,~R. Accurate nucleon-nucleon potential with charge-independence breaking. \emph{Physical Review C} \textbf{1995}, \emph{51}, 38\relax
\mciteBstWouldAddEndPuncttrue
\mciteSetBstMidEndSepPunct{\mcitedefaultmidpunct}
{\mcitedefaultendpunct}{\mcitedefaultseppunct}\relax
\EndOfBibitem
\bibitem[Schatz(1996)]{schatz1996scattering}
Schatz,~G.~C. Scattering theory and dynamics: time-dependent and time-independent methods. \emph{The Journal of Physical Chemistry} \textbf{1996}, \emph{100}, 12839--12847\relax
\mciteBstWouldAddEndPuncttrue
\mciteSetBstMidEndSepPunct{\mcitedefaultmidpunct}
{\mcitedefaultendpunct}{\mcitedefaultseppunct}\relax
\EndOfBibitem
\bibitem[Wang(2000)]{wang2000several}
Wang,~D. \emph{Several new developments in quantum reaction dynamics}; New York University, 2000\relax
\mciteBstWouldAddEndPuncttrue
\mciteSetBstMidEndSepPunct{\mcitedefaultmidpunct}
{\mcitedefaultendpunct}{\mcitedefaultseppunct}\relax
\EndOfBibitem
\bibitem[Manolopoulos \latin{et~al.}(1993)Manolopoulos, Stark, Werner, Arnold, Bradforth, and Neumark]{manolopoulos1993transition}
Manolopoulos,~D.~E.; Stark,~K.; Werner,~H.-J.; Arnold,~D.~W.; Bradforth,~S.~E.; Neumark,~D.~M. The transition state of the F+ H2 reaction. \emph{Science} \textbf{1993}, \emph{262}, 1852--1855\relax
\mciteBstWouldAddEndPuncttrue
\mciteSetBstMidEndSepPunct{\mcitedefaultmidpunct}
{\mcitedefaultendpunct}{\mcitedefaultseppunct}\relax
\EndOfBibitem
\bibitem[Weeks and Tannor(1994)Weeks, and Tannor]{weeks1994time}
Weeks,~D.~E.; Tannor,~D.~J. A time-dependent formulation of the scattering matrix for the collinear reaction H+ H2 (v) -> H2(v)+ H. \emph{Chemical physics letters} \textbf{1994}, \emph{224}, 451--458\relax
\mciteBstWouldAddEndPuncttrue
\mciteSetBstMidEndSepPunct{\mcitedefaultmidpunct}
{\mcitedefaultendpunct}{\mcitedefaultseppunct}\relax
\EndOfBibitem
\bibitem[BAE(1982)]{bae1982review}
BAE,~M. A review of quantum-mechanical approximate treatments of three-body reactive systems. \textbf{1982}, \relax
\mciteBstWouldAddEndPunctfalse
\mciteSetBstMidEndSepPunct{\mcitedefaultmidpunct}
{}{\mcitedefaultseppunct}\relax
\EndOfBibitem
\bibitem[Siegbahn(1978)]{liu1978LSTH}
Siegbahn,~B.,~P.;~Liu An Accurate Three‐dimensional Potential Energy Surface for H3. \emph{The Journal of Chemical Physics} \textbf{1978}, \emph{68}, 2457–2465\relax
\mciteBstWouldAddEndPuncttrue
\mciteSetBstMidEndSepPunct{\mcitedefaultmidpunct}
{\mcitedefaultendpunct}{\mcitedefaultseppunct}\relax
\EndOfBibitem
\bibitem[Truhlar and Horowitz(1978)Truhlar, and Horowitz]{truhlar1978functional}
Truhlar,~D.~G.; Horowitz,~C.~J. Functional representation of Liu and Siegbahn’s accurate ab initio potential energy calculations for H+ H2. \emph{The Journal of Chemical Physics} \textbf{1978}, \emph{68}, 2466--2476\relax
\mciteBstWouldAddEndPuncttrue
\mciteSetBstMidEndSepPunct{\mcitedefaultmidpunct}
{\mcitedefaultendpunct}{\mcitedefaultseppunct}\relax
\EndOfBibitem
\bibitem[Frishman \latin{et~al.}(1997)Frishman, Hoffman, and Kouri]{frishman1997distributed}
Frishman,~A.; Hoffman,~D.~K.; Kouri,~D.~J. Distributed approximating functional fit of the H 3 ab initio potential-energy data of Liu and Siegbahn. \emph{The Journal of chemical physics} \textbf{1997}, \emph{107}, 804--811\relax
\mciteBstWouldAddEndPuncttrue
\mciteSetBstMidEndSepPunct{\mcitedefaultmidpunct}
{\mcitedefaultendpunct}{\mcitedefaultseppunct}\relax
\EndOfBibitem
\bibitem[Schatz and Kuppermann(1973)Schatz, and Kuppermann]{schatz1973role}
Schatz,~G.~C.; Kuppermann,~A. Role of direct and resonant (compound state) processes and of their interferences in the quantum dynamics of the collinear H+ H2 exchange reaction. \emph{The Journal of Chemical Physics} \textbf{1973}, \emph{59}, 964--965\relax
\mciteBstWouldAddEndPuncttrue
\mciteSetBstMidEndSepPunct{\mcitedefaultmidpunct}
{\mcitedefaultendpunct}{\mcitedefaultseppunct}\relax
\EndOfBibitem
\bibitem[Chernoff(1981)]{chernoff1981note}
Chernoff,~H. A note on an inequality involving the normal distribution. \emph{The Annals of Probability} \textbf{1981}, 533--535\relax
\mciteBstWouldAddEndPuncttrue
\mciteSetBstMidEndSepPunct{\mcitedefaultmidpunct}
{\mcitedefaultendpunct}{\mcitedefaultseppunct}\relax
\EndOfBibitem
\bibitem[Childs \latin{et~al.}(2021)Childs, Su, Tran, Wiebe, and Zhu]{childs2021theory}
Childs,~A.~M.; Su,~Y.; Tran,~M.~C.; Wiebe,~N.; Zhu,~S. Theory of trotter error with commutator scaling. \emph{Physical Review X} \textbf{2021}, \emph{11}, 011020\relax
\mciteBstWouldAddEndPuncttrue
\mciteSetBstMidEndSepPunct{\mcitedefaultmidpunct}
{\mcitedefaultendpunct}{\mcitedefaultseppunct}\relax
\EndOfBibitem
\end{mcitethebibliography}


\end{document}